\documentclass[12pt]{iopart}
\usepackage{graphicx}
\usepackage{iopams}  
\usepackage[square, comma, sort&compress, sectionbib,numbers]{natbib}

\def\newblock{\hskip .11em plus .33em minus .07em}

\begin{document}

\title[Probing weak localization in graphene with a movable scatterer]{Imaging coherent transport in graphene (Part II): Probing weak localization}

\author{Jesse Berezovsky and Robert M Westervelt}

\address{School of Engineering and Applied Science, and Department of Physics, \\Harvard University, Cambridge MA 02138.}
\ead{westervelt@seas.harvard.edu}

\begin{abstract}
Graphene has opened new avenues of research in quantum transport, with potential applications for coherent electronics.  Coherent transport depends sensitively on scattering from microscopic disorder present in graphene samples: electron waves traveling along different paths interfere, changing the total conductance.  Weak localization is produced by the coherent backscattering of waves, while universal conductance fluctuations are created by summing over all paths. In this work, we obtain conductance images of weak localization with a liquid-He-cooled scanning probe microscope, by using the tip to create a movable scatterer in a graphene device.  This technique allows us to investigate coherent transport with a probe of size comparable to the electron wavelength.  Images of magnetoconductance \textit{vs.} tip position map the effects of disorder by moving a single scatterer, revealing how electron interference is modified by the tip perturbation.  The weak localization dip in conductivity at $B=0$ is obtained by averaging magnetoconductance traces at different positions of the tip-created scatterer.  The width $\Delta B_{WL}$ of the dip yields an estimate of the electron coherence length $L_\phi$ at fixed charge density.  This ``scanning scatterer'' method provides a new way of investigating coherent transport in graphene by directly perturbing the disorder configuration that creates these interferometric effects.
\end{abstract}

\maketitle
\section{Introduction}
\begin{figure}[htbp]
\centering
\includegraphics[width=.75\textwidth]{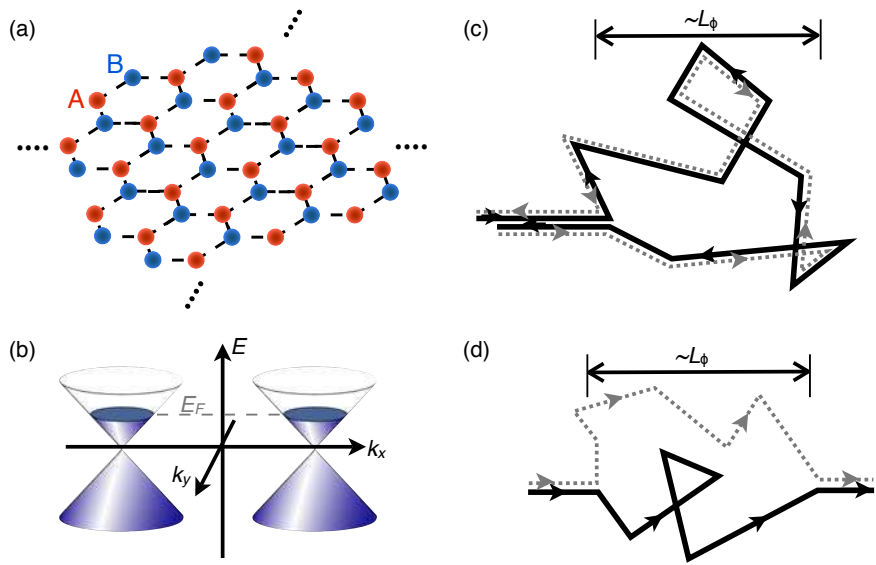}
\caption{(a) The hexagonal graphene lattice with carbon atoms in the two sublattices labeled A (red) and B (blue).  (b) Schematic diagram of graphene's band structure showing how the two pairs of conical conduction and valence bands each meet at a point in $k$-space - the Dirac point.  The states are filled up to the Fermi energy $E_F$. (c) Illustration of a pair of time-reversed backscattered trajectories that interfere to cause weak localization.  The diffusive coherence length $L_\phi$ bounds the set of trajectories that can interfere. (d) A pair of forward scattering trajectories whose interference contributes to universal conductance fluctuations (UCF).  \label{fig:theory}} 
\end{figure}

Graphene, a single atomic layer of carbon atoms in a hexagonal lattice (figure~\ref{fig:theory}a), shows quantum phenomena~\cite{Geim:2007} from the coherent flow of electron waves.  These include the quantum Hall effect~\cite{Zhang:2005b}, the Josephson effect~\cite{Heersche:2007} and weak localization~\cite{Tikhonenko:2008}.  Graphene has an unusual band structure created by the interaction of electrons on the two sublattices, labeled A and B in  (figure~\ref{fig:theory}a): the conduction and valence bands are conical, and they meet at a point in $k$-space, the Dirac point, like the band structure for a massless relativistic particle. The ability to tune the charge density in graphene from positive values for electrons through zero to negative densities for holes, with no energy gap, means that the Fermi wavelength can be comparable to the mean free path, and the charge density can be broken up into puddles of electrons and holes by disorder.  

Coherent electron waves traveling through disordered conductors show weak localization~\cite{Altshuler:1980,Bergmann:1984,Beenakker:1991} and universal conduction fluctuations (UCF) ~\cite{Lee:1985, Altshuler:1985,Washburn:1986}; both effects are created by the interference of electron waves.  Weak localization results from the coherent backscattering of waves from a disordered potential (figure~\ref{fig:theory}c), while UCF are created by the interference of waves traveling along all possible paths (figure~\ref{fig:theory}d).  The set of paths that interfere is bounded by the diffusive coherent length $\sim L_\phi$.  Because of the unique band structure of graphene, unique behavior for weak localization and UCF has been predicted~\cite{Suzuura:2002,McCann:2006, Rycerz:2007,Kharitonov:2008,Kechedzhi:2009}, and experiments are beginning to shed light on these issues~\cite{Tikhonenko:2008,Berezovsky:2009,Morozov:2006,Berger:2006,Graf:2007,Wu:2007,Ki:2008,staley:2008}. 

In this work, we use a liquid-He-cooled scanning probe microscope to probe weak localization in graphene by mapping the effect of a single SPM-tip-created scatterer on coherent electron transport. While much can be learned from bulk transport measurements, a nanoscale probe that perturbs the system on the same size scale as the disorder potential and the electron wavelength provides important new information~\cite{Tessmer:1998,Topinka:2000,Topinka:2001,Topinka:2003,Pioda:2004,Jura:2007,Martin:2008,Braun:2008}.  By scanning the location of the scatterer created by the SPM tip, we spatially map the change $\Delta G$ in conductance $G$ caused by the tip, producing a conductance image $\mathbb{G}(n,B)$ that represents a ``fingerprint'' of the intrinsic disorder, at a particular density $n$ and magnetic field $B$.  The magnitude of the conductance fluctuations is ${\sim e^2/h}$, their lateral size is $\sim 10$s of nm, comparable to the Fermi wavelength, and the images repeat, as found for UCF conductance images of graphene in zero applied magnetic field~\cite{Berezovsky:2009}. The conductance images change with magnetic field, and become uncorrelated when $\Delta B$ is larger than a characteristic correlation field $\Delta B_c$.  Weak localization is observed as a dip in the magnetoconductance $G$ at $B=0$, with a characteristic width $\Delta B_{WL}$.  We study weak localization by averaging over either a range of back gate voltages $V_g$ to vary the charge density, or over a set of different tip positions at a fixed $V_g$. We find predicted agreement between the values of $\Delta B_{WL}$ for weak localization and $\Delta B_c$ for UCF and, as discussed below.  Both $\Delta B_c$ and $\Delta B_{WL}$ increase as the charge density $n$ is reduced, and display a maximum at the Dirac point.

\section{Theory}

Graphene has an unusual bandstructure.  A graphene sheet is composed of a single layer of carbon atoms in a hexagonal lattice, with two atoms in each unit cell forming sublattices labeled A and B in figure~\ref{fig:theory}a.  The band structure has two valleys: each valley has a conduction and a valence band that meet at a single point in the Brillouin zone, known as the Dirac point, with no energy gap (figure~\ref{fig:theory}b).  The electron and hole energies increase linearly with momentum $|\mathbf{k}|$ near their meeting point, and they are nominally isotropic, yielding pairs of conical bands that are symmetric about the Dirac point.  Because of the two atoms in each unit cell, there is an additional degeneracy in the electronic states of graphene known as chirality, with the eigenstates of opposite chirality arising from the two equivalent sublattices.  Scattering between the two valleys in graphene and the effects of chirality must be considered in understanding weak localization in graphene, as discussed below.

The phenomenon of weak localization can be understood by considering interference between time-reversed paths of backscattered electrons in a disordered material~\cite{Altshuler:1980,Bergmann:1984}.  Figure~\ref{fig:theory}c illustrates a pair of time-reversed paths that contribute to backscattering: both paths have an incoming wavevector $\mathbf{k_{in}}$, and an outgoing wavevector $\mathbf{k_{out}}=-\mathbf{k_{in}}$.  The only difference between them is that one traverses the loop in a clockwise direction, and the other traverses the loop in a counterclockwise direction.  As an electron travels along each path, it accumulates phase according to the Dirac equation in the case of graphene, or the Schr\"odinger equation in more conventional materials.  This phase is controlled by the integral of the electron momentum and the magnetic potential along the path, as well as the geometric phase acquired in graphene.  In conventional materials such as metals and semiconductors with no spin-orbit coupling, the geometric phase difference between these paths around a closed loop is zero, yielding constructive interference at $B=0$ that enhances backscattering, causing a net reduction of the conductance (hence, weak localization).  In graphene, the accumulation of geometric phase depends on whether elastic scattering breaks the chiral symmetry of the electron state or scatters electrons between valleys; these processes lead to both constructive and destructive contributions to the interference, as discussed below.  

An applied magnetic field acts to destroy weak localization by altering the phase of electrons traveling in opposite directions along time-reversed paths.  In a perpendicular magnetic field $B$, a phase shift $\delta\phi = \pm 2\pi A B/\Phi_0$ is picked up in both paths in figure~\ref{fig:theory}c given by the magnetic flux penetrating the effective area $A$ enclosed by the diffusive loop, where $\Phi_0=h/e$ is the magnetic flux quantum - the sign depends on the direction of propagation around the loop.  As the magnetic field is increased, the interference present at $B=0$ is destroyed, allowing one to observe weak localization experimentally as a dip in magnetoconductance at $B=0$, with width $\Delta B_{WL}$ determined by the average effective area $A$ enclosed by pairs of backscattered paths.  

In the discussion above, we have assumed that the transport is entirely coherent, whereas realistically, inelastic scattering also occurs that randomizes the phase of an electron wave.  If inelastic scattering occurs at a rate $1/\tau_\phi$ and we assume an elastic scattering rate $1/\tau_e\gg1/\tau_\phi$, then we can define a diffusive coherence length $L_\phi = (D\tau_\phi)^{1/2}$, where $D$ is the electron diffusion constant; $L_\phi$ is the average distance over which an electron remains coherent.  The coherence length $L_\phi$ sets an upper limit to the size of diffusive loops that give rise to weak localization (see figures~\ref{fig:theory}c and \ref{fig:theory}d), resulting in a characteristic magnetic field 
\begin{equation}
\Delta B_{WL} \approx \frac{h}{2 e L_\phi^2} 
 \label{eq:WLwidth}
 \end{equation}
required to destroy weak localization, by producing a $2\pi$ phase shift between time-reversed paths in an area $L_\phi^2$. 

The conductivity correction $\Delta g_{WL}(B)$ due to weak localization is calculated by using the diffusion equation for coherent electron transport.  In a two-dimensional conventional metal or semiconductor the change $\Delta g$ in conductivity $g$ \textit{vs.} magnetic field $B$ is~\cite{Altshuler:1980}
\begin{equation}
\Delta g_{WL} = \frac{\gamma e^2}{2\pi h}F\left(\frac{4eB}{\hbar}L_\phi^2\right)
\label{eq:WLmetal}
\end{equation}
where $\gamma$ is the total spin and valley degeneracy, and $F(z) = \ln(z) + \psi(1/2 + 1/z)$, with $\psi(x)$ the digamma function.  Equation~\ref{eq:WLmetal} is calculated by considering the diffusion equation in the limit where the sample dimensions $L,W \gg L_\phi$, and the elastic mean free path $l_e \ll l_\phi$.  In graphene, equation~\ref{eq:WLmetal} becomes~\cite{McCann:2006}
\begin{eqnarray}
\Delta g = \frac{e^2}{\pi h}&\left[ F\left( \frac{4eB}{\hbar}L_\phi^2\right) - F\left( \frac{4eB}{\hbar(L_\phi^{-2}+2L_i^{-2})}\right) \right.-\nonumber\\
 &\left.- 2 F\left( \frac{4eB}{\hbar(L_\phi^{-2}+L_i^{-2}+L_*^{-2})}\right) \right] ,
\label{eq:WLgraphene}
\end{eqnarray} 
where $\tau_*^{-1} =DL_*^{-2}$ is the rate of chirality-breaking scattering within a valley, and $\tau_i^{-1} =DL_i^{-2}$ is the rate of scattering between valleys.

Because equation~\ref{eq:WLgraphene} has both positive and negative terms, the conductivity correction in graphene may be positive or negative at $B=0$ depending on the relative rates of the different scattering processes.  A positive, weak anti-localization correction to the conductivity has been observed in epitaxially-grown graphene samples~\cite{Wu:2007}.  In exfoliated graphene samples, however, such as the ones considered here, the negative weak localization correction is typically dominant at low temperatures~\cite{Heersche:2007,Tikhonenko:2008,Ki:2008}, implying the existence of a significant source of chirality-breaking scattering, possibly from lattice defects in the graphene.  We observe a negative weak localization conductivity correction in all samples studied, and obtain qualitatively good fits to equation~\ref{eq:WLgraphene} with fit parameters similar to those in Ref.~\cite{Tikhonenko:2008}.  However, equation~\ref{eq:WLgraphene} is derived for $L,W \gg L_\phi$, which is not the case for our samples.  In samples A and B (see figure~\ref{fig:samples}b), $W\sim L_\phi$, and in sample C, we have $W\sim L_\phi$ near the leads and $W\gg L_\phi$ in the central region.  For simplicity, we will use equation~\ref{eq:WLwidth} to estimate $L_\phi$ from measurements of $\Delta B_{WL}$ below.

In addition to weak localization, the interference of electron waves traveling along different paths creates universal conductance fluctuations (UCF) with standard deviation $\delta G \sim e^2/h$ independent of sample size or the degree of disorder~\cite{Lee:1985, Altshuler:1985}. This coherent correction to the conductivity depends sensitively on the positions of the scatterers.  We have recently shown~\cite{Berezovsky:2009} that moving a single scatterer created by an SPM tip by a distance comparable to the electron wavelength is sufficient to cause the full range of conductance fluctuations in zero magnetic field, in agreement with theoretical predictions~\cite{Feng:1986, Altshuler:1985b}. 

For UCF at finite B, the change $\Delta B_c$ in magnetic field needed to reduce the correlation of the conductance by a factor of two is~\cite{Ferry:1999}   
\begin{equation}
\Delta B_{c}\approx\frac{h}{e L_\phi^2}. 
\label{eq:UCFwidth}
\end{equation}
Note that $\Delta B_{c}=2 \Delta B_{WL}$ because weak localization arises from the phase difference accumulated along two counter-propagating paths around a diffusive loop, whereas UCF comes from a single loop. We compare these theoretical predictions with our experimental results below.

\section{Experimental Methods}
\begin{figure}[tbp]
\centering
\includegraphics[width=.45\textwidth]{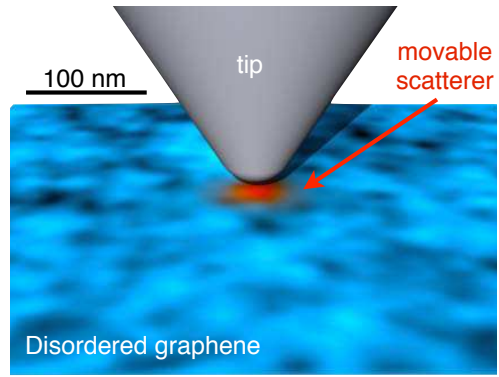}
\caption{Schematic diagram (to scale) of the charged scanning probe microscope tip creating an additional scatterer (red) in graphene.  The shaded blue color scale in the graphene layer represents the disorder potential.   \label{fig:schematic}}
\end{figure}

In this work, we use a liquid-He-cooled SPM tip to create a movable scatterer in a mesoscopic graphene Hall bar.  As shown schematically in figure~\ref{fig:schematic}, a voltage-biased SPM tip is brought into close proximity with a graphene sample.  Via capacitive coupling, the tip creates an image charge as a local change in the graphene charge density, indicated by the red spot, that adds to the average charge density $n$ controlled by a back gate voltage $V_g$, and to the existing disorder in the graphene layer, shown in blue, which consists of randomly placed scatterers created by charged impurities located above or below the graphene sheet~\cite{rossi:2008,Zhang:2009}.  As discussed below, the effect of the tip is to add one additional scatterer in the sample which can be moved about at will.  

\begin{figure}[bp]
\centering
\includegraphics[width=.35\textwidth]{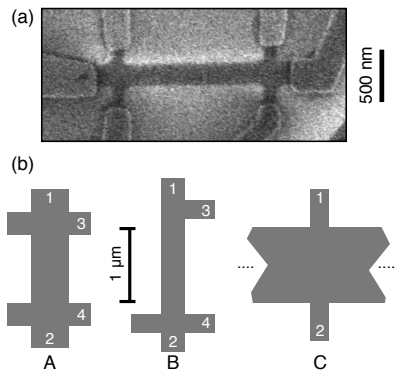}
\caption{(a) Scanning electron micrograph of a graphene Hall bar, contacted by six Cr/Au leads. (b) Schematic diagrams of the graphene samples measured here (to scale).  Sample C continues to the left and right for several microns.  An ac current is applied between contacts 1 and 2 for all three samples, and the voltage is measured between contacts 3 and 4 for samples A and B, and between contacts 1 and 2 for sample C. \label{fig:samples}}
\end{figure}

A scanning electron micrograph of a graphene sample is shown in figure~\ref{fig:samples}a. The samples studied in these experiments are fabricated from single-layer graphene flakes deposited by mechanical exfoliation (the ``sticky tape method'') on a degenerately doped Si substrate capped with 280~nm of SiO$_2$.  The samples were contacted by Cr/Au leads defined by electron beam lithography, and the graphene structures were formed by a mask defined by electron beam lithography followed by an oxygen plasma etch.  A planar back gate voltage $V_g$ is applied between the sample and the conducting Si substrate, allowing us to tune the charge density $n$ in the graphene.  

The data presented here are from three graphene samples, shown schematically in figure~\ref{fig:samples}b; all showed the same qualitative behavior.  Samples A and B were measured in a four-probe geometry, with lead 2 grounded, and a root-mean-square (rms) current $I= 25$~nA at 5 kHz applied between leads 1 and 2; the voltage for conductance measurements was measured between leads 3 and 4, with a lock-in amplifier.  Sample C was tested using a two-probe geometry with the same current applied between leads 1 and 2, with the voltage also measured across leads 1 and 2.  For sample C, we estimate the contact resistance $R_0$ by finding the value of $R_0$ that fits the characteristic conductance expression $G=1/(R-R_0)$ \textit{vs.} $V_g$, which is linear on either side of the Dirac point $V_{Dirac}$; similar data is shown below in figure~\ref{fig:sweeps}a for the four-probe geometry.  We find $R_0=7500~\Omega$, which we subtract from the data from sample C.

Each sample is mounted in a home-built scanning probe microscope~\cite{Topinka:2000, Aidala:2007}, and cooled in He exchange gas that is in thermal contact with a liquid He bath at $T=4.2$~K.  The sample sits in the core of a superconducting magnet that provides a magnetic field up to $B=6$~T perpendicular to the sample plane.  We verify that the samples are single-layer graphene by observing quantum Hall conductance plateaux at the expected values of  $4(\nu+1/2)e^2/h$, where $\nu$ is an integer.  For the scanning probe measurements, a conducting SPM tip with radius of curvature $r_{tip} = 20$~nm is held at a constant height $h_{tip} = 10$~nm above the graphene sample.  The tip is grounded, so the tip charge is set by the contact potential between the graphene and the degenerately doped Si tip.  Charged impurities on the surface of the graphene layer create image charges in the tip, which also contribute to the tip charge.

Using transport measurements and electrostatic simulations, we characterized our graphene samples, and determined the spatial profile of the density perturbation created by the SPM tip.  Classical electrostatic simulations (Maxwell, Ansoft LLC) were used to determine the charge density profile in the graphene created by the back gate and the SPM tip.  The back gate voltage $V_g$ capacitively creates a uniform charge density $n=\alpha(V_g-V_{Dirac})$, with $\alpha = 8\times10^{10}$~V$^{-1}$cm$^{-2}$, where $V_{Dirac}$ is the offset of the Dirac point from $V_g=0$ caused by charged impurities near the graphene.  The mobility $\mu$ for samples A, B, and C was $\mu_A=7200$~cm$^2/$Vs, $\mu_B=5600$~cm$^2/$Vs, and $\mu_C=4200$~cm$^2/$Vs, found from the measured $G$ \textit{vs.} $V_g$ curves.  The local perturbation to the graphene charge density caused by the tip has a Lorentzian-like shape with half-width at half maximum (HWHM) $\sim 25$~nm, and a maximum $\sim 5\times 10^{11}$~cm$^{-2}$ for a potential difference $\sim1$~V between the tip and the sample.  Previous experiments using scanning tunneling microscopy~\cite{Deshpande:2009,Zhang:2009} and a scanning single-electron-transistor charge sensor~\cite{Martin:2008} have observed the disorder in graphene samples on a SiO$_2$ substrate to consist of charge puddles with peak charge density $\sim 4\times10^{11}$~cm$^{-2}$ and lateral dimensions $l \sim 20$~nm.  The perturbation created by the tip in this paper has approximately the same amplitude as these naturally occurring charge density fluctuations, and about twice the width.  Therefore, it is reasonable to think that the tip adds an additional, controllable scatterer to the pre-existing scatterers created by charged defects above and below the graphene. 

\begin{figure}[htbp]
\centering
\includegraphics[width=\textwidth]{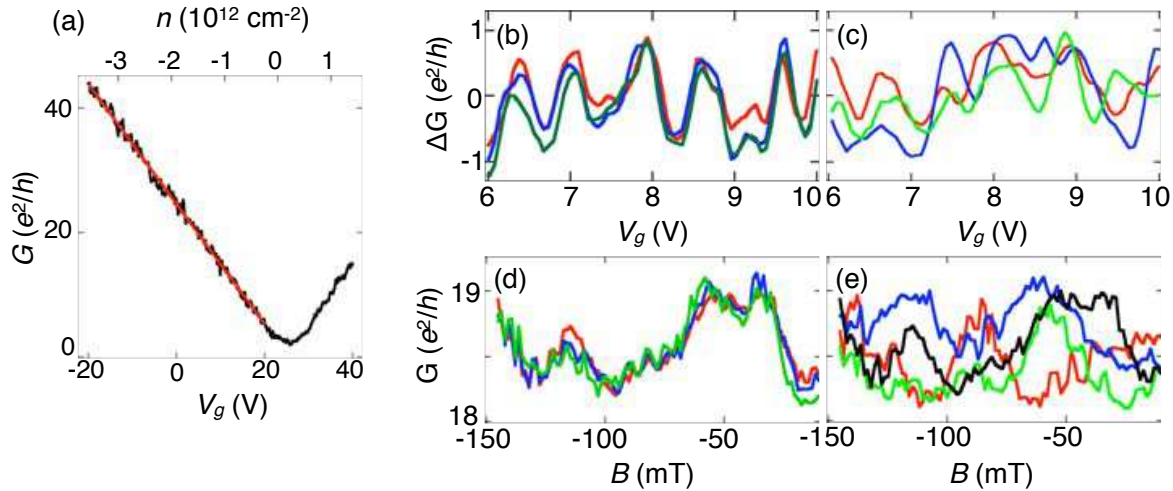}
\caption{(a) Conductance $G$ of graphene sample A measured in a four-probe geometry \textit{vs.} back gate voltage $V_g$ at $B=0$.  Red line is a linear fit of $G$ \textit{vs.} $V_g$ below 20~V. (b) Fluctuations $\Delta G$ \textit{vs.} $V_g$ about the linear fit ($B=0$); the traces repeat when the SPM tip is at a fixed position far from the sample. (c) Same as (b), but with the tip at height $h_{tip}=10$~nm above the sample for three different lateral positions spaced $100$~nm apart, showing a clear dependence of the conductance fluctuations on tip position.  (d) Magnetoconductance traces $G$ \textit{vs.} $B$ with $V_g=0$~V ($n=-1.8\times10^{12}$~cm$^{-2}$) and $h_{tip}=10$~nm, at three tip positions spaced 5~nm apart; the traces repeat. (e) Same as (d), with four tip positions $100$~nm apart; the traces now differ.\label{fig:sweeps}}
\end{figure}

\section{Experimental Results}

Using an SPM tip, we probe the dependence of coherent magnetotransport on the position of a single scatterer.  Figure~\ref{fig:sweeps}a shows the conductance $G$ of sample A \textit{vs.} $V_g$ at zero magnetic field; the data vary linearly with $V_g$ on either side of the Dirac point $V_{Dirac}=22$~V as indicated by the linear fit shown in red, as expected for single atomic layer graphene.  The small fluctuations $\Delta G$ away from the linear fit are reproducible, and are attributed to UCF~\cite{Berezovsky:2009}.  The fluctuations $\Delta G$ can be seen more clearly by subtracting the linear background from $G$, as is shown over a small range of $V_g$ in figure~\ref{fig:sweeps}b.  The three traces of $\Delta G$ \textit{vs.} $V_g$ repeat the same scan with the SPM tip fixed far from the sample (tip height $h_{tip}>100~\mu$m), demonstrating good reproducibility.  The rms amplitude $\delta G\sim e^2/h$ is in agreement theory~\cite{Lee:1985, Altshuler:1985}.  Figure~\ref{fig:sweeps}c shows the effect of moving the charged SPM tip to three different positions spaced $100$~nm apart, at $h_{tip}=10$~nm above the sample -- the traces are now quite different.  These data demonstrate that the motion of a single scatterer created by the SPM tip is sufficient to rearrange the conductance fluctuation pattern.

Moving the tip also changes fluctuations in the magnetoconductance $G$.  Figure~\ref{fig:sweeps}d shows three traces of $G$ \textit{vs.} $B$ with $V_g=0$ ($n=-1.8\times10^{12}$~cm$^{-2}$) and with the tip at three nearby positions, only 5~nm apart.  Here the fluctuations repeat, with amplitude $\delta G \sim e^2/h$.  In contrast, when the tip is moved to several locations spaced $100$~nm apart (figure~\ref{fig:sweeps}e), the magnetoconductance is largely uncorrelated and looks like a different sample.  Clearly, there is a critical length $>5$~nm beyond which the tip must be moved in order to decorrelate the magnetoconductance fluctuations.

\subsection{Conductance images and correlations in a magnetic field}

\begin{figure}[htb]
\centering
\includegraphics[width=.85\textwidth]{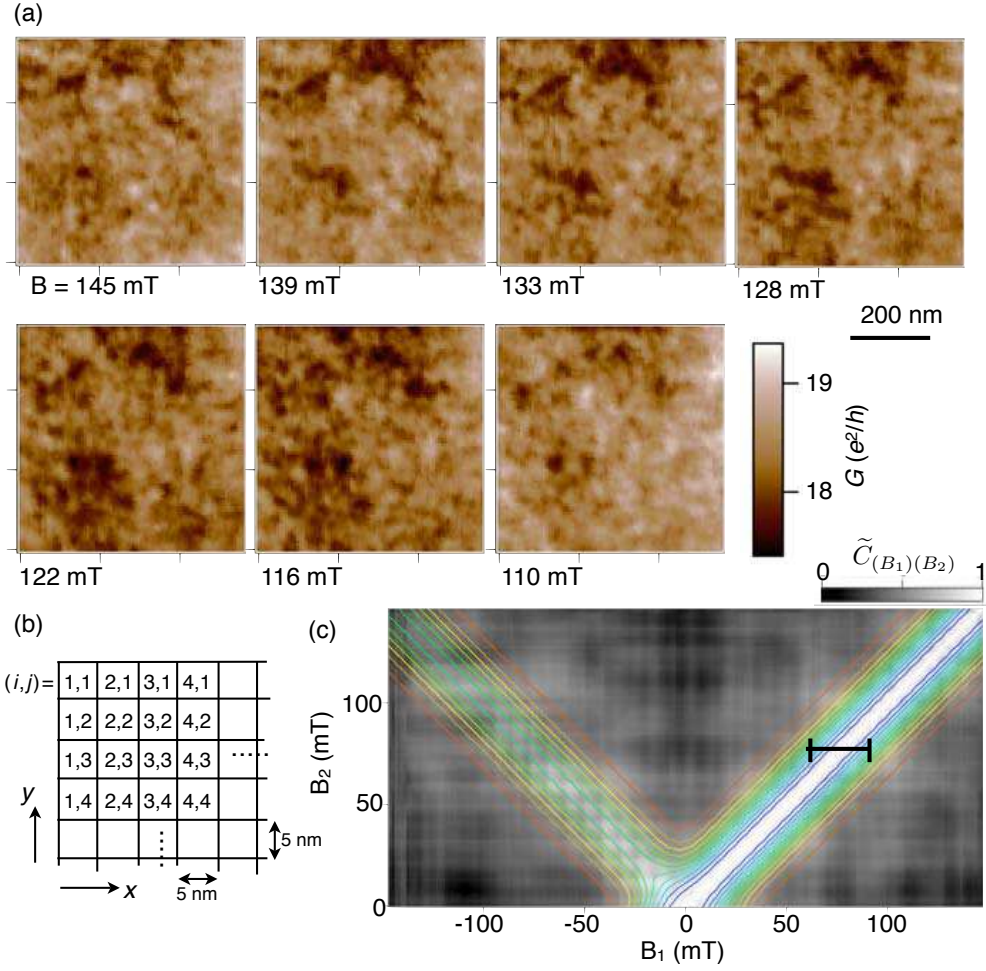}
\caption{(a) A series of conductance images $\mathbb{G}(B,V_g)$ \textit{vs.} magnetic field $B$ at fixed back gate voltage $V_g=0$~V (sample A).  (b) Schematic of a conductance image $\mathbb{G}$, with pixels spaced 5~nm apart displaying the conductance $G_{\mathbf{r}_{ij}}$ with the tip located at position $\mathbf{r}_{i,j}=(x_i,y_j)$.  (c) Correlation $\widetilde{C}_{(B_1)(B_2)}$ between conductance images $\mathbb{G}(B_1,V_g)$ and $\mathbb{G}(B_2,V_g)$ with $V_g=0$~V.  Lines show contours of a best fit from which $\Delta B_c$ is obtained; the width of the black bar is 2$\Delta B_c$. \label{fig:crosscorr}}
\end{figure}

We obtain conductance images $\mathbb{G}(B,V_g)$ by recording $G$ as the tip is raster scanning over a $400\times 400$~nm$^2$ region in the center of the sample, at fixed magnetic field $B$ and back gate voltage $V_g$.  A conductance image $\mathbb{G}$ consists of $80\times80$ pixels that display the sample conductance $G(\mathbf{r}) = G_{\mathbf{r}_{i,j}}$ over a two-dimensional array of tip positions $\mathbf{r}_{i,j} = (x_i,y_j)$ with pixel spacing $\Delta x = \Delta y = 5$~nm, as shown schematically in figure~\ref{fig:crosscorr}b.  The standard deviation $\delta G_{tip}$ of the conductance $G$ in an image is given by the standard deviation of $G_{\mathbf{r}_{ij}}$ over all values of $i$ and $j$.  By acquiring a series of images $\mathbb{G}(B,V_g)$ as $B$ or $V_g$ is stepped to different values, we study the effect of the tip on coherent transport.  

A series of conductance images $\mathbb{G}(B,V_g)$ \textit{vs.} $B$ is shown in figure~\ref{fig:crosscorr}a with $V_g = 0$ ($n=-1.8\times10^{12}$~cm$^{-2}$). The conductance images show fluctuations $\Delta G$ \textit{vs.} $\mathbf{r}$ with standard deviation $\delta G_{tip} \sim e^2/h$ and lateral sizes  $\sim 10$s of nm comparable to the Fermi wavelength and to the size of the scatterer created by the tip.  The conductance images $\mathbb{G}(B,V_g)$ repeat over time intervals $\sim 1$~hr. as expected for coherent conductance fluctuations, demonstrating that their origin is not temporal noise.
It can be seen by eye that the series of conductance images $\mathbb{G}(B,V_g)$ shown in figure~\ref{fig:crosscorr}a change with $B$ in a continuous fashion.  We quantify this change by calculating the cross-correlation $C_{AB}=\int (G_A(\mathbf{r})-\langle G_A \rangle)(G_B(\mathbf{r})-\langle G_B \rangle)d\mathbf{r}$ between conductance images $\mathbb{G}_A=G_A(\mathbf{r}_{i,j})$ and $\mathbb{G}_B=G_B(\mathbf{r}_{i,j})$, where angle brackets denote the average over the tip position $\mathbf{r}$.  We then define a normalized correlation $\widetilde{C}_{AB}$, such that the autocorrelation of an image with itself is unity
\begin{equation}
 \widetilde{C}_{AB} = \frac{C_{AB}}{(C_{AA}C_{BB})^{1/2}}.
 \label{eq:crosscorr}
\end{equation}

The grayscale map in figure~\ref{fig:crosscorr}c displays correlations $\widetilde{C}_{(B_1)(B_2)}$ between images $\mathbb{G}(B_1)$ and $\mathbb{G}(B_2)$ taken from a series of 200 images spanning $B=\pm145$~mT  at $V_g = 0$ (as in figure~\ref{fig:crosscorr}a).  By definition, $\widetilde{C}_{(B)(B)}=1$.  The fainter peak along $B_1=-B_2$ shows that there is some symmetry in the magnetoconductance fluctuations as the sign of $B$ is reversed about $B=0$.  The symmetry is imperfect because of the nature of a coherent four-probe measurement~\cite{Ferry:1999}, and because of slow drift of the images over the course of the measurement time $\simeq 10$~hrs.  We fit the map in figure~\ref{fig:crosscorr}c to a function given by the sum of two Gaussians centered at $B_1=\pm B_2$, shown as the contour plot in figure~\ref{fig:crosscorr}c.   From this fit, we extract the magnetic correlation length $\Delta B_{c}=17$~mT, seen in figure~\ref{fig:crosscorr}c to be constant over the range of $B$ shown here; $\Delta B_c$ represents the magnetic field change needed to reduce the correlation between conductance images by one half.  Using equation~\ref{eq:UCFwidth}, we find the diffusive coherence length $L_{\phi c} = 500$~nm, where $L_{\phi c}$ represents a value of $L_\phi$ obtained from $\Delta B_c$.  A series of images $\mathbb{G}(B,V_g)$ \textit{vs.} $B$ provides a means to obtain $\Delta B_c$ and $L_{\phi c}$ at a fixed back gate voltage $V_g$, which we will return to below to study the behavior of $\Delta B_c$ \textit{vs.} $V_g$.  

The amplitude $\delta G_{tip}$ of conductance fluctuations created by moving the local perturbation of the SPM tip differs from the amplitude $\delta G_{V_g}$ created by changing the the global back gate voltage $V_g$.  This is understood by considering the ratio $L_\phi/L$ of the diffusive coherence length $L_\phi$ to the sample length $L$~\cite{Beenakker:1991}.  A sample with width $W\sim L_\phi$ and length $L > L_\phi$ can be envisioned as $N=L/L_\phi$ uncorrelated, coherent regions in series, each with conductance $G_i = G_0$ and exhibiting coherent conductance fluctuations with standard deviation $\delta G_i$.  The total conductance $G$ of the sample is 
\begin{equation}
G = \left( \sum_i \frac{1}{G_i} \right)^{-1} = G_0/N.
\label{eq:sumG}
\end{equation}
For a global change such as a change in $V_g$, all $N$ regions undergo uncorrelated  conductance fluctuations with standard deviation $\delta G_i = \delta G_0$.  The standard deviation $\delta G_{V_g}$ in the total conductance is given by taking partial derivatives of equation~\ref{eq:sumG} summed in quadrature:
\begin{equation}
\delta G_{V_g} = \left[ \sum_i \left( \frac{\partial G}{\partial G_i} \delta G_i \right)^2 \right]^{1/2}
= N^{-3/2}\delta G_0
= (L_\phi/L)^{3/2}\delta G_0.
\label{eq:global}
\end{equation}   
Alternatively, the perturbation from the SPM tip acts locally, causing conductance fluctuations in only one coherent region yielding $\delta G_{j} = \delta G_0$ and $\delta G_i = 0$ for $i \neq j$.  For a local perturbation, the standard deviation $\delta G_{tip}$ in the total conductance is
\begin{equation}
\delta G_{tip} = \left[ \sum_i \left( \frac{\partial G}{\partial G_i} \delta G_i \right)^2 \right]^{1/2}
= N^{-2}\delta G_0
= (L_\phi/L)^{2}\delta G_0.
\label{eq:local}
\end{equation}   
Combining equations~\ref{eq:global} and \ref{eq:local} we obtain $\delta G_{tip}/\delta G_{V_g}=(L_\phi/L)^{1/2}$.

We obtain the standard deviations $\delta G_{tip}$ and $\delta G_{V_g}$ experimentally from conductance images $\mathbb{G}$ and traces of $G$ \textit{vs.} $V_g$ for sample A.  To reduce the uncertainty, we average $\delta G_{tip}$ over 20 conductance images with back gate voltage stepped from $V_g = -1$ to 1~V at $B=0$ yielding $\delta G_{tip}=0.32~e^2/h$.  We calculate $\delta G_{V_g}=0.50~e^2/h$ from the standard deviation of conductance traces $G$ \textit{vs.} $V_g$ in the same range of $V_g = -1$ to 1~V, measured with the tip away from the sample and with the linear background subtracted, as in figure~\ref{fig:sweeps}b.  We find that the fluctuation $\delta G_{tip}$ created by the tip is smaller than that from the back gate voltage by the ratio $\delta G_{tip}/\delta G_{V_g}=0.64$.  This value is in good agreement with theory, which predicts $\delta G_{tip}/\delta G_{V_g}=(L_\phi/L)^{1/2}=0.6$, using $L_{\phi c}= 500$~nm obtained above at $V_g=0$.

\subsection{Weak localization}

\begin{figure}[htbp]
\centering
\includegraphics[width=.5\textwidth]{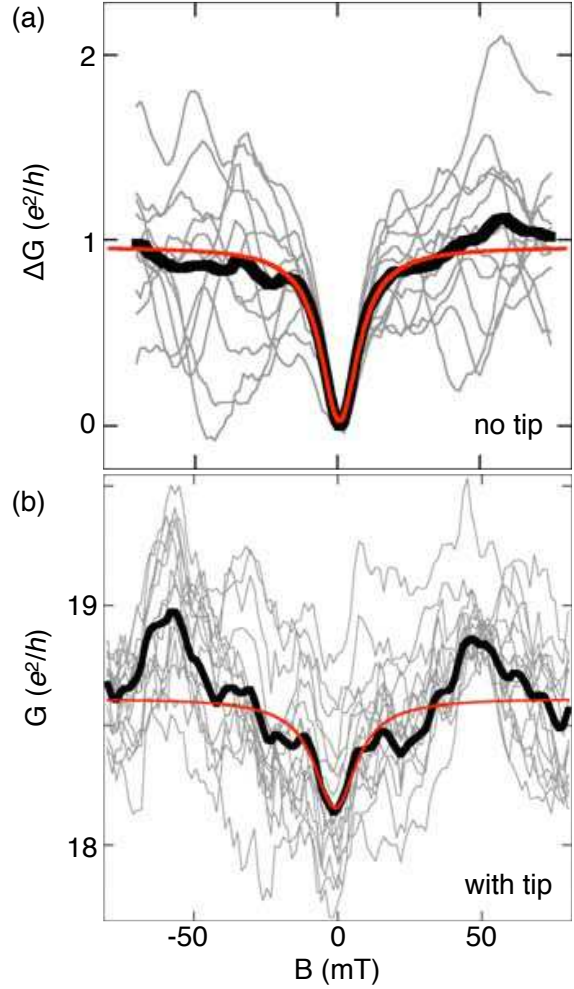}
\caption{(a) Weak localization conductance dip in sample A measured with the SPM tip fixed far from the sample. Gray: Magnetoconductance $\Delta G$ \textit{vs.} $B$ at 11 densities stepped from $n=-1.6\times 10^{12}$~cm$^{-2}$ to $n=-2.0\times 10^{12}$~cm$^{-2}$; curves are shifted to $\Delta G=0$ at $B=0$.  Black curve: Average of the 11 gray curves. Red: Fit of black curve to a Lorentzian, yielding the width $\Delta B_{WL}=8.1\pm0.5$~mT. (b) Weak localization dip in sample A measured at tip height $h_{tip}=10$~nm.  Gray: Magnetoconductance $G$ \textit{vs.} $B$ at a set of different tip positions separated by 100 nm, with fixed density $n=-1.8\times 10^{12}$~cm$^{-2}$.  Black curve: Average of 6400 such curves with tip positions spanning a 5~nm grid. Red: Fit of black curve to a Lorentzian, yielding width $\Delta B_{WL}=9\pm2$~mT.\label{fig:WL}}
\end{figure}

Weak localization can be identified by the conductance dip (or peak) that occurs at zero magnetic field.  A conductance dip $\Delta G_{WL}$ created by weak localization is shown in the magnetoconductance traces in figure~\ref{fig:WL}. As for other measurements of exfoliated graphene~\cite{Heersche:2007,Tikhonenko:2008,Ki:2008}, we observe a conductance dip from weak localization, not a peak.  In our samples, where the sample size is comparable to $L_\phi$, the conductance change $\Delta G_{WL}$ from weak localization is comparable to that from UCF.  To observe the weak localization effect alone, we average over multiple magnetoconductance traces, by varying either the density $n$ or the tip position $\mathbf{r}$ to average out UCF.  

We clearly observe the weak localization dip at $B=0$ in figure~\ref{fig:WL}a, by averaging magnetoconductance traces at different densities $n$ with the tip away from the sample ($h_{tip} > 100~\mu$m).  The gray lines in  figure~\ref{fig:WL}a show a series of 11 magnetoconductance traces for densities $n$ stepped by $\Delta n = 4 \times 10^{10}$~cm$^{-2}$ between $n=-1.6\times 10^{12}$~cm$^{-2}$ and $n=-2.0\times 10^{12}$~cm$^{-2}$, a range $\Delta V_g = 5$~V.  Because $G$ changes significantly with $n$, these traces have been shifted to pass through the same point at $B=0$.  A weak localization dip $\Delta G_{WL}$ is clearly seen in the average of all 11 traces, shown by the black curve, with magnetoconductance fluctuations averaged out. 

We also observe the weak localization conductance dip $\Delta G_{WL}$ in figure~\ref{fig:WL}b by averaging magnetoconductance traces $G_{\mathbf{r}_{ij}}$ \textit{vs.} $B$ over a grid of different tip positions $\mathbf{r}_{ij}$.  The gray lines show magnetoconductance traces at fixed $n=-1.8\times 10^{12}$~cm$^{-2}$ and different tip positions, on a grid spaced by $\Delta x = \Delta y =100$~nm, with $h_{tip}=10$~nm.  This data is obtained from a series of 200 conductance images $\mathbb{G}(B,V_g)$ at different magnetic fields $B$, stepped from  $B=-145$~mT to $B=+145$~mT.  A magnetoconductance trace $G_{\mathbf{r}_{i,j}}(B)$ at tip position $\mathbf{r}_{ij}$ is extracted from the dataset by taking the same pixel from each image $\mathbb{G}(B,V_g)$ at different values of $B$.  The weak localization conductance dip $\Delta G_{WL}$ is visible in the black curve in figure~\ref{fig:WL}b, which is the average of all 6400 traces.  Conductance fluctuations are still clearly visible in the averaged magnetoconductance, because the tip-created scatterer only affects the conductance in a region of size $\sim L_\phi$, about one third of the sample.  Magnetoconductance fluctuations arising from regions of the sample at distances $>L_\phi$ from the tip are not decorrelated by changing the tip position.  

By measuring the width $\Delta B_{WL}$ of the weak localization conductance dip $\Delta G_{WL}$ shown in figures~\ref{fig:WL}a and \ref{fig:WL}b, we obtain an estimate for the coherence length $L_{\phi WL}$.  To obtain $\Delta B_{WL}$, we fit the averaged curves of $G$ \textit{vs.} $B$ in figures~\ref{fig:WL}a and \ref{fig:WL}b to a Lorentzian function, which approximates the predicted lineshapes given in equations~\ref{eq:WLmetal} and ~\ref{eq:WLgraphene} for weak localization in two dimensions~\cite{Beenakker:1991,McCann:2006}.  The red lines in figures~\ref{fig:WL}a and \ref{fig:WL}b show the results of the Lorentzian fit, with HWHM $\Delta B_{WL}=8.1\pm0.5$~mT and $\Delta B_{WL}=9\pm2$~mT in figures~\ref{fig:WL}a and b respectively, in good agreement with each other.  Using equation~\ref{eq:WLwidth}, we obtain a coherence length $L_{\phi WL} =510\pm20$~nm for figure~\ref{fig:WL}a and $L_{\phi WL} =480\pm50$~nm for figure~\ref{fig:WL}b.  These values obtained from weak localization measurements match the estimate $L_{\phi c}= 500$~nm obtained above from cross-correlations between magnetoconductance images shown in figure~\ref{fig:crosscorr}.

\begin{figure}[htbp]
\centering
\includegraphics[width=.5\textwidth]{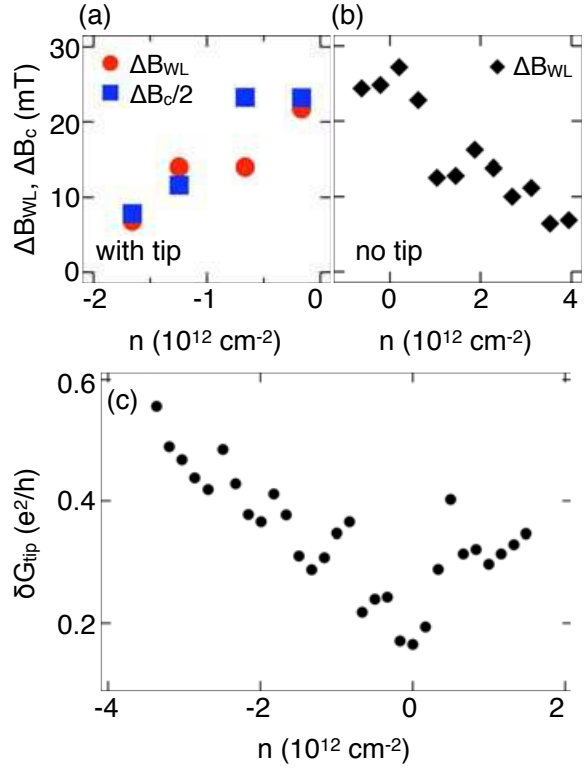}
\caption{(a) Widths of the weak localization conductance dip $\Delta B_{WL}$ and the correlation magnetic field $\Delta B_c$ for conductance fluctuations \textit{vs.} charge density $n$ in sample C. Circles: $\Delta B_{WL}$ \textit{vs.} $n$ obtained from magnetoconductance traces averaged over different tip positions, as in figure~\ref{fig:WL}b. Squares: $\Delta B_c/2$ obtained from cross-correlations between conductance images, as in figure~\ref{fig:crosscorr}c. (b) $\Delta B_{WL}$ in sample B, obtained with the tip fixed far from the sample, by averaging magnetoconductance traces over a range ($4\times 10^{11}$~cm$^{-2}$) of $n$, showing qualitative agreement with the data in (a).  (c) Standard deviation $\delta G_{tip}$ of the conductance $G_{\mathbf{r}_{ij}}$ over all tip positions $\mathbf{r}_{ij}$ in a conductance image $\mathbb{G}$ \textit{vs.} $n$ for sample A.\label{fig:lphi}}
\end{figure}

Measurements of $\Delta B_{WL}$ from the weak localization conductance dip $\Delta G_{WL}$ and measurements of  $\Delta B_c$ from cross-correlations between conductance images allow us to study how the phase coherence length $L_{\phi}$ depends on the charge density $n$. Previous bulk weak localization measurements found the coherence length $L_{\phi WL}$ in graphene varies significantly with $n$, with a minimum at the Dirac point~\cite{Tikhonenko:2008,Ki:2008}. Figure~\ref{fig:lphi}a shows our data for $\Delta B_{WL}$ and $\Delta B_{c}/2$ \textit{vs.} $n$, obtained from conductance images $\mathbb{G}(B,V_g)$ for sample C. The agreement clearly shows the factor of 2 difference, predicted by the interference of two time-reversed diffusive loops for weak localization \textit{vs.} the single loop for interfering forward scattering paths, discussed above.  Each data point is derived from a series of 200 conductance images $\mathbb{G}(B,V_g)$ at a fixed density $n$. The circles show $\Delta B_{WL}$, from the width of the weak localization conductance dip in an average over tip positions, as in figure~\ref{fig:WL}b.  The squares show $\Delta B_{c}/2$ obtained from the cross-correlation $\widetilde{C}_{(B_1)(B_2)}$ (equation~\ref{eq:crosscorr}) between conductance images $\mathbb{G}(B_1,V_g)$ and $\mathbb{G}(B_2,V_g)$.  We find $\Delta B_c$ from the correlation function $\widetilde{C}_{(B_1)(B_2)}$ as in figure~\ref{fig:crosscorr}c. The values of $\Delta B_{WL}$ and $\Delta B_{c}$ in figure~\ref{fig:lphi}a both increase as the hole density decreases and the system approaches the Dirac point.  

Figure~\ref{fig:lphi}b shows $\Delta B_{WL}$ found for sample B with the tip fixed far from the sample.  Each data point is obtained by averaging 20 magnetoconductance traces $G$ \textit{vs.} $B$ over a range of densities $n=\pm 2\times10^{11}$~cm$^{-2}$ (a 5V change in $V_g$). The data in figure~\ref{fig:lphi}b show that $\Delta B_{WL}$ increases as the electron density is decreased, and is largest at the Dirac point ($n=0$), complementing the behavior for holes.

The coherence length $L_\phi$ for samples B and C is estimated from $\Delta B_{WL}$ and $\Delta B_{c}$ in figures~\ref{fig:lphi}a and \ref{fig:lphi}b, using equations \ref{eq:WLwidth} and \ref{eq:UCFwidth}.  We find that the minimum coherence lengths $L_{\phi WL}=L_{\phi c}= 300$~nm occur at the Dirac point for both samples.  The maximum coherence length $L_\phi$ occurs at the largest hole or electron density: $L_{\phi WL}=600$~nm for $n=4\times10^{12}$~cm$^{-2}$ in sample B, and $L_{\phi WL}=L_{\phi c}= 500$~nm at $n=-1.7\times10^{12}$~cm$^{-2}$ in sample C.  The change of $L_\phi$ with $n$ is in good agreement with previous results~\cite{Tikhonenko:2008,Ki:2008}, and is likely caused by the change in electron-electron interaction strength with $n$~\cite{Morozov:2006, Tikhonenko:2008,Ki:2008}, as well as the effects of electron-hole puddles near the Dirac point~\cite{staley:2008, Tikhonenko:2008,Ki:2008}.  

The standard deviation $\delta G_{tip}$ of conductance fluctuations over a single image for a series of conductance images $\mathbb{G}(n)$ at different densities $n$ at $B=0$ is shown in figure~\ref{fig:lphi}c.  We find that  $\delta G_{tip}$ increases away from the Dirac point with the electron or hole density; these results are similar to bulk transport measurements on single and multilayer graphene~\cite{Graf:2007,staley:2008,Horsell:2009}.  The trend in $\delta G_{tip}$ \textit{vs.} $n$ is consistent with the behavior of $\Delta B_{WL}$ and $\Delta B_c$ in figures~\ref{fig:lphi}a and \ref{fig:lphi}b.  As discussed above, when the sample length $L>L_\phi$, then $\delta G_{tip} \propto (L_\phi/L)^2$.  Because $\Delta B_{WL}\propto L_{\phi WL}^{-2}$ and $\Delta B_{c} \propto L_{\phi c}^{-2}$, we expect $\delta G_{tip} \propto 1/\Delta B_{WL}$ and $\delta G_{tip} \propto 1/\Delta B_c$; the data in figure~\ref{fig:lphi} follow this trend.  From $n=0$ to $n=\pm 2\times10^{12}$~cm$^{-2}$, $\delta G_{tip}$ changes by a factor $\approx 2$, which agrees with the observed change in $\Delta B_{WL}$ and $\Delta B_{c}$ by a factor $\approx 1/2$ over the same range of $n$ in figures~\ref{fig:lphi}a and \ref{fig:lphi}b. 

\section{Conclusions}
The study of coherent electron transport and weak localization in graphene has recently received much attention through bulk transport measurements.  Here, we add a new technique that allows one to probe transport by creating a movable scatterer with an SPM tip.  This allows us to investigate weak localization in a new way by creating a controllable, local change to the disorder configuration. Our measurements provide a direct way to observe the spatial nature of coherent transport in graphene.  This technique provides an image of the magnetoconductance \textit{vs.} tip position that represents a spatial ``fingerprint'' of the interfering paths at a particular Fermi energy $E_F$ and magnetic field $B$.  From correlations between images at different $B$ we measure the coherence length $L_{\phi c}$ for electrons diffusing through the sample, and find good agreement with the coherence length $L_{\phi WL}$ obtained from weak localization measurements.

Understanding and controlling disorder is one of the main challenges in realizing many of the proposals for probing new physics and discovering applications for graphene.  In this work, we have described a tool that maps the effect of a nanoscale change to the disorder in a graphene sample produced by an SPM tip.  This represents a step towards the goal of gaining mastery over disorder in graphene, by developing methods to controllably shape it and bend it to our will.

\ack
We thank K. Brown, M. F. Borunda, and E. J. Heller for helpful discussions, and acknowledge support from the Department of Energy under grant DE-FG02-07ER46422.

\end{document}